# Uniting the order and disorder dynamics in photoexcited VO$_2$

Hao-Wen Liu[1,2,§], Wen-Hao Liu[1,2,§], Zhao-Jun Suo[1,2], Zhi Wang[1], Jun-Wei Luo[1,2]*, Shu-Shen Li[1,2], and Lin-Wang Wang[3]*

[1]*State Key Laboratory of Superlattices and Microstructures, Institute of Semiconductors, Chinese Academy of Sciences, Beijing 100083, China*

[2]*Center of Materials Science and Optoelectronics Engineering, University of Chinese Academy of Sciences, Beijing 100049, China*

[3]*Materials Science Division, Lawrence Berkeley National Laboratory, Berkeley, California 94720, United States*

[§]These authors contributed equally: Hao-Wen Liu, Wen-Hao Liu.
*Email: jwluo@semi.ac.cn; lwwang@lbl.gov

**Abstract:**

Photoinduced phase transition (PIPT) is always treated as a coherent process, but ultrafast disordering in PIPT is observed in recent experiments. Utilizing the real-time time-dependent density functional theory (rt-TDDFT) method, here, we track the motion of individual vanadium (V) ions during PIPT in VO$_2$ and uncover that their coherent or disordered dynamics can be manipulated by tuning the laser fluence. We find that the photoexcited holes generate a force on each V-V dimer to drive their collective coherent motion, in competing with the thermal-induced vibrations. If the laser fluence is so weak that the photoexcited hole density is too low to drive the phase transition alone, the PIPT is a disordered process due to the interference of thermal phonons. We also reveal that the photoexcited holes populated by the V-V dimerized bonding states will become saturated if the laser fluence is too strong, limiting the timescale of photoinduced phase transition.

**Introduction:**

Photoexcitation using ultrafast laser pulses provides a powerful approach to manipulate the material properties with a timescale in the limit of atomic motion [1-9]. It was believed that the atomic motion in the photoinduced phase transitions (PIPTs) is in collective coherent dynamics [10-13]. But, recent experiments show an ultrafast disordering of atomic motions in PIPTs of $VO_2$ [14] and $Rb_{0.3}MoO_3$ [15], challenging the conventional view of phase transitions. As an archetypical 3d-correlated oxide [16,17], manipulating the transition from the monoclinic ($M_1$) insulator phase to rutile (R) metallic phase in $VO_2$, as shown in Fig. 1a and 1d, is a popular topic with numerous efforts that have focused on elucidating the evolution of both the electronic and lattice degrees of freedom. Photoinduced insulator-to-metal transition (IMT) is ultrafast with sub-femtosecond timescales [18,19]. Using a four-dimensional (4D) femtosecond electron diffraction to measure the evolution of the V-V dimers following the photoexcitation, P. Baum *et al.* [20] discovered that the motion of the V atoms is first along the direction of the V-V bond (*a* axis) with femtosecond timescales, then along the *b* and *c* axis within picosecond timescales, thus claiming a coherent structural transition from $M_1$ to R phase. Besides, some experimental groups have reported that the PIPT in $VO_2$ originates from a coherent motion of V-V dimers at 6 THz phonon mode [10,21-23]. However, in a recent experiment, S. Wall *et al.* [14] used a femtosecond total x-ray scattering method beyond the general x-ray or electron diffraction [2,12,20,24-26] to measure averages over many unit cells and they discovered the motion of V-V dimers is disordered in PIPT rather than a collective motion along the coherent phonon coordinate.

Besides the debate of atomic ordering or disordering in the phase transition, the timescale of $M_1$-to-R phase transition in $VO_2$ also has a major controversy. Early experiments reported the timescale of PIPT being gradually reduced from 100 fs to 40 fs [18] with increasing laser fluence [10,18,19]. However, M. R. Otto *et al.* [25] recently observed far longer timescales of 200-500 fs in the lacking of obvious relation with the laser fluence. This has been attributed to disordered movements of the atoms, and M. R. Otto *et al.* supported that the PIPT in $VO_2$ should be viewed as a disordering or even melting transition [25]. How to consolidate these different experimental claims, both in order and disorder phase transition and in their corresponding timescales, is thus a major challenge in this field.

In this work, to investigate the coherent or disordered manner of atomic motion, we have utilized our newly developed rt-TDDFT algorithm [6,27,28] to simulate the PIPT in $VO_2$. We show that the phase transition is in atomic disordering with a timescale of about 200 fs at low laser fluence and, at high laser fluence, becomes a coherent manner with a fluence-dependent timescale below 100 fs. The disordering phase transition is driven by a combination of the atomic driving forces caused by photoexcitation and thermal phonon vibration. On the other hand, the coherent phase transition is only controlled by the photoexcited atomic driving forces, and the thermal phonon is only a small perturbation. We also demonstrate that, with increasing the laser fluence, the smallest timescale of phase transition is indeed saturated to a minimum value of ~ 55 fs, which is due to the saturation of the population of the V-V bonding states by the photoexcited holes. Our findings provide a unified theory for understanding the atomic coherent and disordering motion in PIPT, uniting different experimental results.

**Results:**

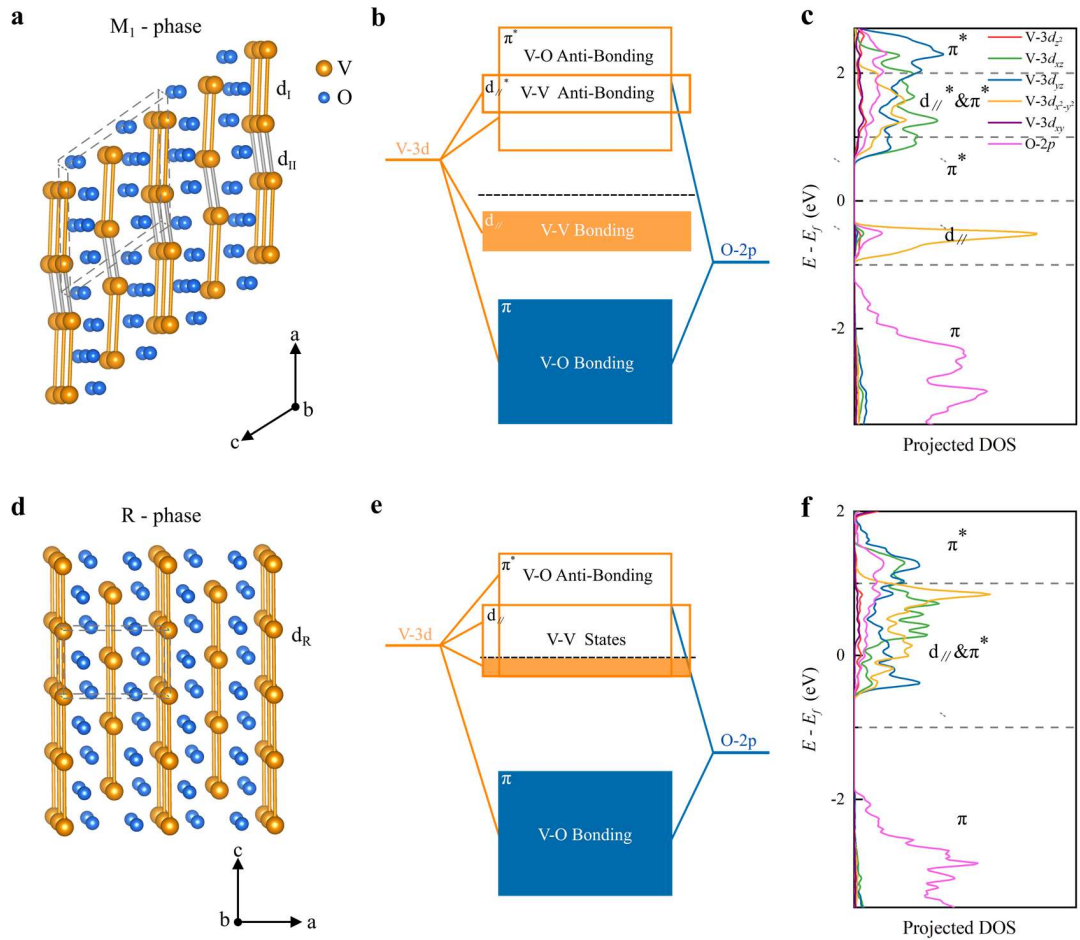

**Figure 1. VO$_2$ atomic and electronic structures. a,** The 2×2×2 supercell structure of M$_1$-phase VO$_2$ in which $d_I$ and $d_{II}$ represent the V-V long and short bonds. Gray parallelepipeds represent the unit cell of the VO$_2$ structures. V and O atoms are labeled as orange and blue balls, respectively. **b,** Schematic of the VO$_2$ bonding and antibonding states of M$_1$-phase VO$_2$. **c,** The projected density of states (PDOS) of M$_1$-phase VO$_2$. **d,** The 2×2×3 supercell structure of R-phase VO$_2$ in which all V-V bonds ($d_R$) are equal. **e,** Schematic of the VO$_2$ bonding and antibonding states of R-phase VO$_2$. **f,** The projected density of states (PDOS) of R-phase VO$_2$.

**Atomic structures and electronic orbital properties.** In high temperature, the VO$_2$ is stabilized in a high-symmetry R phase (P4$_2$/mnm) with a vanadium atom surrounded by six oxygen atoms forming an octahedron. When the temperature is below the transition temperature $T_c \sim 340$ K [29,30], the vanadium atoms deviate from the octahedral geometric center to form a low-symmetry M$_1$ phase (P2$_1$/C). The V-O octahedral environment and the coupling between the O $2p$ orbitals with V $3d$ orbitals split the V-3d orbitals into a combination of low-energy triply degenerate $t_{2g}$ states ($d_{x^2-y^2}, d_{xz}$ and $d_{yz}$) and high-energy doubly degenerate $e_g^\sigma$ states ($d_{z^2}$ and $d_{xy}$) [31,32]. The $t_{2g}$ states are further separated into an $a_{1g}$ orbital ($d_{x^2-y^2}$) and two $e_g^\pi$ orbitals ($d_{xz}$ and $d_{yz}$) because of the V-O octahedral structure with different V-O distances (which breaks the cubic symmetry). Notably, the $a_{1g}$ orbital is parallel to the rutile $c$ axis ($c_R$), which hardly hybridizes with O $2p$ orbitals to form V-O bonds. The dimerization of the V atoms leads to twisted V-V pairs, splitting the highly directional $a_{1g}$ orbital into a bonding state ($d_\parallel$) and an anti-bonding state ($d_\parallel^*$) (Fig. 1b and 1c). The antibonding orbital energy is raised over the original orbital energy by an amount V (usually referred to as the overlap parameter), the same as the bonding orbital energy decrease [33]. Manipulating the electrons on the $d_\parallel$ bonding state to occupy the $d_\parallel^*$ antiboding state by photoexcitation can break the V-V dimers, inducing a M$_1$-to-R phase transition.

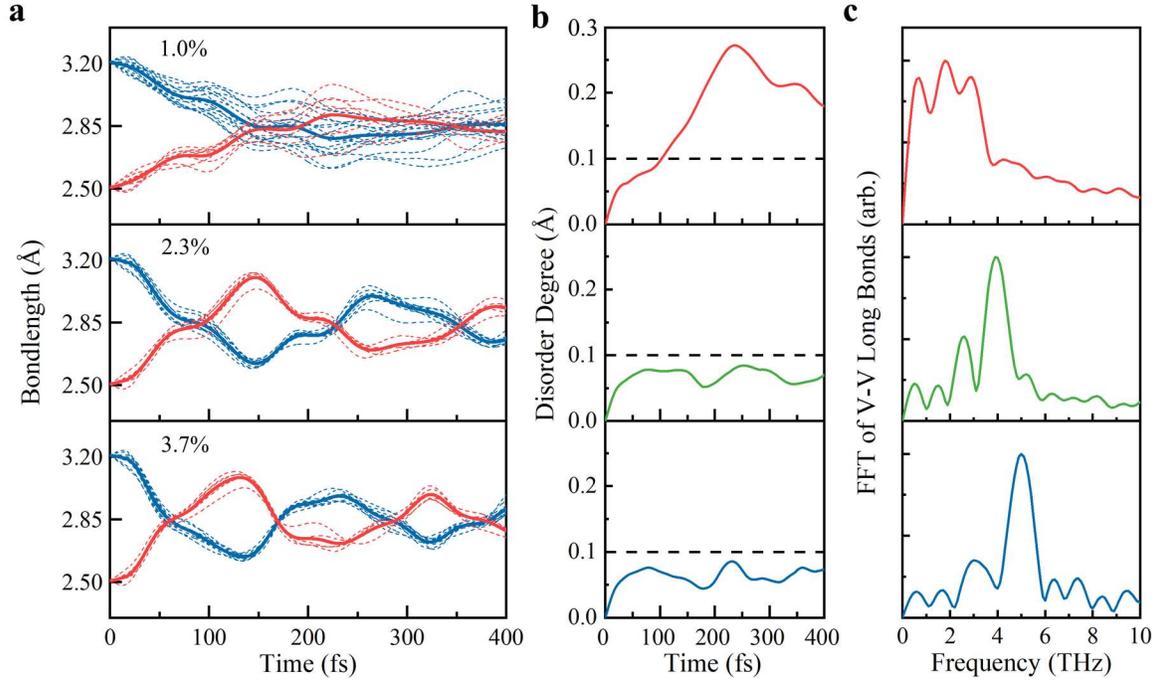

**Figure 2. Photoinduced ultrafast phase transition in different electronic excitations. a,** The bond-length evolution under the different laser pumping. The blue and red dotted lines represent the bond lengths of each long and short bond, respectively. The blue and red solid lines represent the average bond length of long and short bonds. **b,** The disorder degree evolutions at different excitations. **c,** Phonon modes of V atoms during the phase transition, which are obtained from the normalized fast Fourier transform (FFT) of the average bond length of V-V bonds.

**Coherence and disordering of atomic motion.** To simulate the photoinduced $M_1$-to-R ultrafast phase transition, we utilize an 800 nm laser to pump electrons from valence bands to conduction bands of $VO_2$ in our rt-TDDFT simulations. The initial lattice temperature is set to 50 K. In photoexcited simulations, we use an external electric field with a Gaussian shape (Supplementary Fig. 1a),

$$E(t) = E_0 cos(\omega t) \exp[-(t-t_0)^2/(2\sigma^2)] \quad (1)$$

to simulate the laser pulse. Here, $t_0$ = 17.5 fs, the pulse width $\sqrt{2}\sigma$ = 7 fs, and the photon energy $\omega$ = 1.55 eV, which are consistent with experimental parameters [14,25] (see Method). Besides, we use $E_0$ to tune the laser fluence. With increasing $E_0$ from 0.1 to 0.5, the photoexcited electrons from the valence band to conduction band gradually increase from 1.0% to 6.3% of valence electrons (Supplementary Fig. 1b).

We show the evolution of V-V bonds to track the $M_1$-to-R phase transition in Fig. 2a and Fig. S2a. In the $M_1$ phase, V atoms deviate from the oxygen octahedral center so that V atoms have two types of V-V bonds, including a group of long V-V bonds ($d_I$) and a group of short V-V bonds ($d_{II}$) called V-V dimers. All V-V bonds ($d_R$) have the same value in the R phase. The V-V long bonds ($d_I$ = 3.21 Å) and the V-V short bonds ($d_{II}$ = 2.51 Å) gradually become equal ($d_I = d_{II} = d_R$ = 2.84 Å) with time evolution in our simulations representing the process of photoinduced $M_1$-to-R phase transition, as shown in Fig. 2a and Supplementary Fig. 2a. We discover that there are different types of dynamical processes with different photoexcitation levels. At the low electronic excitations (The percentage of excited electrons $\eta$ = 1.0%), the changes of individual long bonds and individual short bonds during the PIPT are not all the same. Rather, their changes can be characterized as disordered motions, and the phase transition time $\tau$ is nearly 187 fs. With increasing laser fluence, the $d_L$ and $d_S$ can rapidly become the same within 100 fs, and the motions of all bonds are nearly synchronous, hence can be characterized as coherent dynamics.

To describe the degree of disorder in atomic motion, we define a disorder parameter ($L_{disorder}$)

$$L_{disorder} = \sum_{i=1}^{n} \sqrt{(x_i - x_{ave})^2 + (y_i - y_{ave})^2 + (z_i - z_{ave})^2}/n \qquad (2)$$

The summation runs over V atoms, and $n$ represents the number of V-V bonds. The $x_i$, $y_i$, and $z_i$ represent V-V bond length in $x, y, z$ directions. The $x_{ave}$, $y_{ave}$ and $z_{ave}$ represent the average V-V bond length in $x, y, z$ directions. Fig. 2b and Supplementary Fig. 2b shows the $L_{disorder}$ change within the timescale of phase transition in different electronic excitations. At 1.0% electronic excitation, the disordered degree has rapidly reached 0.2 Å before the $M_1$-to-R phase transition. However, the disorder degrees are less than 0.1 Å in other higher excitations. Remarkably, the higher electronic excitation does not increase the disorder at the phase transition point, it decreases it a little bit. We can thus conclude that the electron excitation does not induce disorder. We next proceed with the fast Fourier transform (FFT) for V-V bonds to obtain phonon modes of V atomic motion. Three low-frequency phonon modes from atomic vibrations are discovered at 0.7, 1.8, and 2.9 THz at 1.0% excitation (Fig. 2c). The chaotic phonon modes with three peaks further confirm the disordered process of phase transition at the 1.0% excitation. We only find a single peak between 4-5.5 THz at all higher excitations (Fig. 2c and Supplementary Fig. 2c). These high-frequency modes are generally called coherent phonons in experimental literature [10,21,22], and they are attributed as the cause for the coherent PIPT.

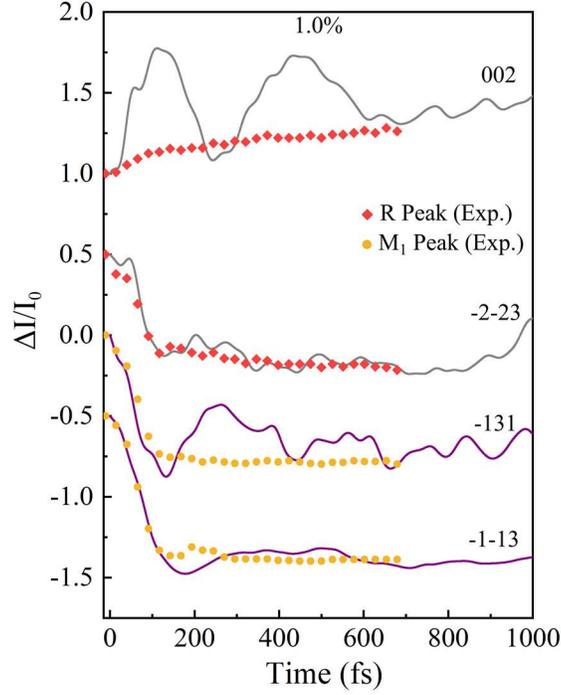

**Figure 3. Time dependence of Bragg peaks for different photoexcitation.** Simulated Bragg peaks [(002)$_R$, (-2-23)$_R$, (-131)$_{M1}$, (-1-13)$_{M1}$] as a function of time is obtained through the structure factors (see Equation 3). Red diamonds and yellow circles show the relative changes of the R-phase peak and $M_1$-phase peak measured by x-ray diffraction experiments.

Furthermore, we obtain the diffraction intensity using the structure factor $F(hkl)$ [14,20],

$$F(h,k,l) = f_V \sum_V \exp[-2\pi i q(hkl) \cdot r_V(t)] + f_O \sum_O \exp[-2\pi i q(hkl) \cdot r_O(t)]$$

$$I(h,k,l) = |F(h,k,l)|^2 \quad (3)$$

Here, $f_V = 23$ and $f_O = 8$ donate the atomic scattering factor of V and O atoms, $q(hkl)$ is the wave-vector, and $r(t)$ is the fractional coordinates of either V or O atoms at time $t$ within the unit cell. The structure vectors $F(h,k,l)$ are calculated at each time step using the atomic positions obtained from the rt-TDDFT results. Subsequently, the diffraction intensities $I(h,k,l)$ are derived from the square of the structure vectors. The final results (Fig. 3 and Supplementary Fig. 3) are the average diffraction intensities over eight unit cells within the 2×2×2 supercell structure. The evolution of Bragg peaks observed here indicates a direct structural transition from the $M_1$-to-R phase. The [(-131)$_{M1}$, (-1-13)$_{M1}$] peaks in the $M_1$ superstructure rapidly drop to zero in intensity, and the peaks labeled [(002)$_R$, (-2-23)$_R$] are present in both phases. These results are in good accord with experimental data [16], as shown in Fig. 3, although the calculated R peak has some oscillations which are absent from the experimental observation.

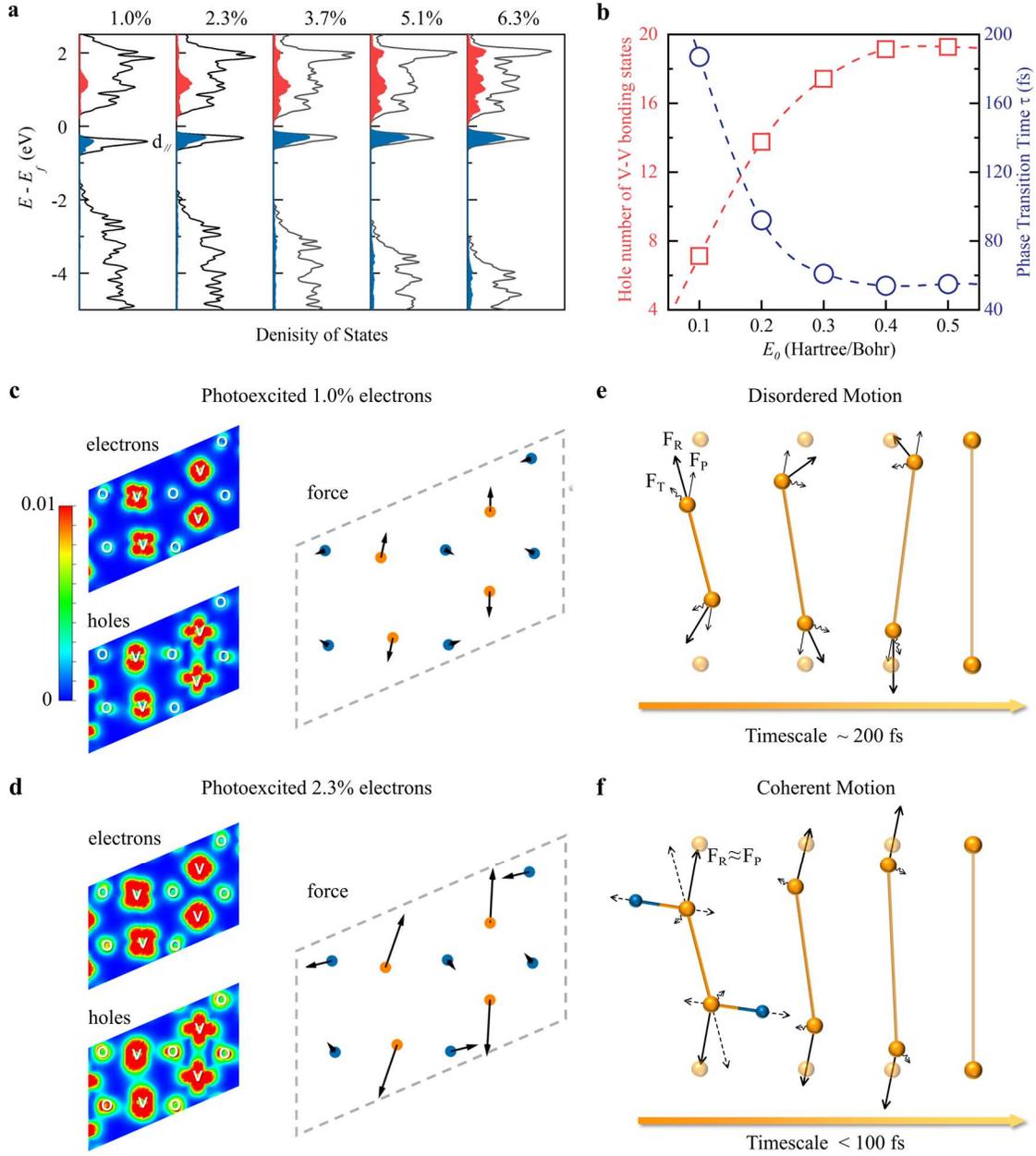

**Figure 4. The atomic microscopic driving force induced by orbital-occupation changes. a,** Density of states (DOS) of photoexcited electrons (red shaded area) and holes (blue shaded area) at different excitations. **b,** The correlation between the holes of V-V bonding states and the phase transition time. The red line shows the hole number of the V-V bonding states excited by different laser fluence. The blue line displays the timescale of $M_1$-to-R phase transition in different laser fluence. **c, d,** Real-space distributions of photoexcited holes, electrons, and driving forces on atoms caused by photoexcitation on $(0\bar{1}1)$ plane at the end of the laser pulses (∼35 fs) for the case of photoexcited 1.0% valence electrons

(c) and the case of photoexcited 2.3% valence electrons (d). **e,** Atomic disordered motion induced by multiple phonons. **f,** Atomic coherent motion caused by strong photoexcitation. The orange balls represent a pair of V-V atoms. The $F_P$ displays the driving force caused by photoexcitation, and the dotted arrows in (f) represent the component of $F_P$ along with the V-V bond and V-O bond directions, respectively. The spiral arrows represent the force from thermal vibrations labeled by $F_T$, and the black arrows show the resultant labeled by $F_R$.

**Microscopic driving force.** Based on the electronic band structure, we show the occupations of photoexcited electrons and holes in Fig. 4a. Valence electrons from the $d_\parallel$ bonding states of V atoms are vertically excited to the $d_\parallel^*$ and $\pi^*$ antibonding states in conduction bands. A strong laser fluence excites more valence electrons, but with similar orbital occupation distributions. Based on our previous theory for IrTe$_2$ [27], if $n$ electrons are excited from bonding states to antibonding states, it will increase the free energy of the system in an amount of about $2nV$, where the overlap parameter $V$ is inversely proportional to the square of the bond length. As a result, to lower the energy, the V-V dimers need to become longer, which yields a driving force to elongate the short V-V bond along the dimer direction. Simultaneously, the occupations of excited electrons on $d_{xz}$ and $d_{yz}$ orbitals lead to a change of bond angles. The total driving force of the V atoms is the result of the combined action of these two forces (Fig. 4f). Furthermore, Figures 4c and 4d show the real-space distributions of photoexcited holes and electrons on $(0\bar{1}1)$ planes at the end of the laser pulses. The photoexcited holes are mainly distributed on V atoms and between V-V dimers, primarily having a V-V bond $d_\parallel$ orbital character. The distributions of photoexcited electrons are more complicated with the occupations on V $d$ ($d_{x^2-y^2}$, $d_{xz}$ and $d_{yz}$) and O $p$ antibonding orbitals, corresponding to V-O antibonding and V-V antibonding characters. Notably, the carrier distributions on V atoms can be divided into two categories with a difference of 90° rotation (Supplementary Fig. 4). Overall, we discover that the photoexcited holes result in a driving force acting on V-V dimers (Fig. 4c and 4d), which can induce V-V dimer dissociation.

The driving force induced by photoexcitation can create a coherent atomic motion. However, we observe a disordering behavior in M$_1$-to-R phase transition in 1.0% excitations. To distinct the different factors, we re-done the rt-TDDFT calculations with the same excitation but at the lattice temperature ($T = 1$ K) to exclude the interference of thermal phonons. In 1.0% excitations, the long bonds (d$_I$) and short bonds (d$_{II}$) scarcely change (no $M_1$-to-R phase transition) within 1.0 ps because the coherent driving forces induced by the lower excitation are not enough to create the phase transition by themselves (Supplementary Fig. 5a). This illustrates that the structural transformation needs the assistance of thermal phonons at a low excitation level. In 2.3% excitations at $T = 1$ K, the long bonds ($d_I$) and short bonds ($d_{II}$) gradually become the same value at 850 fs (Supplementary Fig. 5b), demonstrating that, in this case, the phase transition is dominated by the coherent driving force, instead of the thermal phonon. We thus have two different cases: at lower fluence, the phase transition needs the help of thermal phonon, and the atomic movement appears to be disorder; at higher fluence, the phase transition can be driven by the coherent atomic movement alone, thus the atomic movement appears to be order and coherent.

**Saturation of hole excitations on V-V bonding states.** Our simulations also show the disordered phase transition needs a longer timescale (~ 187 fs), which agrees with the experiment results in disordered PIPT [14,25]. Compared to the disordered phase transition, the PIPT can happen within 100 fs in the coherent dynamics of the V atoms along a straight trajectory from the $M_1$-to-R phase. Interestingly, with increasing electronic excitations, the timescale of phase transition rapidly decreases to a critical value (~ 55 fs) as shown in Fig. 4b, which is also in accordance with the previous experiments [18]. At the same time, the frequency of phonon mode gradually increases and reaches a saturation value (~ 5.5 THz). The bottleneck timescale observed at higher fluence is known as half of the period of the coherent mode in some previous experiments [10,23]. Remarkably, the timescale (~ 55 fs) for phase transition is far lower than the half a period (~ 90 fs) of the 5.5-THz phonon. This phenomenon has also been reported in one experiment [18]. However, the essential reason for the phonon bottleneck and timescale saturation was not illustrated in previous literatures.

For 1.0% and 2.3% excitations, photoexcited holes occupying on V-V bonding states basically. But for 3.7%, 5.1%, and 6.3% excitations, the situations are different. The numbers of total photoexcited holes (or electrons) are 30, 41, and 50, respectively. Whereas the numbers of photoexcited holes occupying on V-V bonding states are 17.4, 19.1, and 19.3 in the simulated supercell (Fig. 4a and 4b). Through the distributions of photoexcited holes projected on $(0\bar{1}1)$ plane (Supplementary Fig. 6), the holes around the O atoms significantly increase with the increase of total hole distributions, while the holes around the V atoms hardly change. The effective hole number on V-V bonding states appears to be saturated, and the rest holes are distributed in V-O bonding states. The saturation phenomenon is caused by the band filling. According to Fermi's "Golden Rule", as the number of electrons on the V-V bonding state decreases, it is more difficult to excite these electrons to the conduction band. At the same time, the higher laser fluence increases the probability of multiphoton absorption of electrons on the deeper valence bands (V-O bonding states) (Fig. 4a). As a result of this, the driving force acting on the V-V dimers is bounded by an upper limit, which determinates the fastest dissociation speed of V-V dimers.

**Discussion**

In the work, we have observed the coherent and disordered phase transition for the $M_1$-to-R phase transition which strongly depends on the number of photoexcited carriers. The disordered process happens within 190 fs in a lower photoexcited density where the thermal phonons are an important factor to make the phase transition possible. During the disordered process, the force direction on vanadium atoms has changed with time due to the effect of thermal phonons (Fig. 4e). The disordered phase transition mechanism and the transition timescale are consistent with the experimental [14,25] and AIMD results within 0.9% electronic excitations [14]. In a higher photoexcitation, the phase transition tends to a coherent dynamic with the coherent phonon mode, and the thermal phonons are only a small perturbation. This is caused by the increase of photoexcited electrons, which produces a larger driving force (Fig. 4f). The timescale of phase transition is also reduced to 100 fs, corresponding to the coherent dynamics reported by the previous experiments of $VO_2$ [10]. With the further increase of photoinduced carriers, the coherent phonon mode gradually arrives at a saturation (~ 5.5 THz) which limits the atomic speed and the timescale of the phase transition (~ 55 fs). Such saturation has been reported experimentally for $VO_2$ [10,18]. Here, we have theoretically illustrated that the phenomenon is from the saturation of hole number on V-V dimers bond states. We believe these phenomena should also exist in other phase-transition materials, such as $TiSe_2$, $TaS_2$, and In atomic wires on Si surfaces [34-36]. Our simulations not only solve the experimental controversies but also provide a powerful view to understand the pathway of ultrafast phase transition by laser-pulse excitations.

**Method:**

**Computational details.** We do the static calculations and rt-TDDFT simulations by the ab-initio package (PWmat) [37]. All calculations are based on the norm-conserving pseudopotentials (NCPP) [38] which are generated by Optimized Norm-Conserving Vanderbilt Pseudopotential and Perdew-Burke-Ernzerhof (PBE) exchange-correlation functional. The wave functions are expanded on a plane-wave basis with an energy cutoff of 50 Ry. V($3s^23p^64s^23d^3$) and O($2s^22p^4$) are treated as valence electrons in the norm-conserving pseudopotentials. In static calculations, a PBE+U exchange-correlation with Hubbard U = 3.4 eV is used. The structure of the $M_1$ phase is fully relaxed by using a conventional unit cell, and an 8×8×8 k-point mesh with Monkhorst-Pack grids is used to sample the Brillouin zone. The resulting lattice constants are $a$ = 5.67Å, $b$ = 4.49 Å and $c$ = 5.32 Å, and beta = 122.47°, in well agreement with experimentally reported lattice constants in the $M_1$ phase at $T$ = 298 K, $a$ = 5.75 Å, $b$ = 4.54 Å, $c$ = 5.38 Å and beta = 122.65° [39].

**TDDFT simulations.** A 2×2×2 supercell (96 atoms), based on the $M_1$ phase fully relaxed unit cell, is used to do rt-TDDFT simulations. In 96-atoms supercell, a 3×3×3 mesh k-point mesh with Monkhorst-Pack grids is used to sample the Brillouin zone. In the rt-TDDFT simulations [28], the time-dependent wave functions $\psi_i(t)$, are expanded by the adiabatic eigenstates $\phi_j(t)$.

$$\psi_i(t) = \sum_l C_{i,l}(t)\phi_l(t) \quad (4)$$

and

$$H(t)\phi_l(t) \equiv \varepsilon_l(t)\phi_l(t) \quad (5)$$

$$H(t) \equiv H(t, R(t), \rho(t)) \quad (6)$$

Here, $R(t)$ represents the nuclear positions and $\rho(t)$ represents the charge density. In equation (4), the evolution of the wave functions $\psi_i(t)$ is changed to the evolution of the coefficient $C_{i,l}(t)$. In equation (5), a linear-time-dependent Hamiltonian (LTDH) is applied to represent the time dependence of the Hamiltonian within a time step. Thus, our rt-TDDFT simulation has a much larger time step (0.1 fs) than the conventional rt-TDDFT (sub-attosecond).

To mimic the photoexcitation, we introduce a uniform A field in reciprocal space [40],

$$H = 1/2(-i\boldsymbol{\nabla} + \boldsymbol{A})^2 = 1/2(-i\nabla_x + A_x)^2 + 1/2(-i\nabla_y + A_y)^2 + 1/2(-i\nabla_z + A_z)^2 \quad (7)$$

The time-space external field can be described as a Gaussian shape in equation (1).

**REFERENCES AND NOTE**

**Acknowledgments**

The work in China was supported by the Key Research Program of Frontier Sciences，CAS under Grant No. ZDBS-LY-JSC019, CAS Project for Young Scientists in Basic Research under Grant No. YSBR-026, the Strategic Priority Research Program of the Chinese Academy of Sciences under Grant No. XDB43020000, and the National Natural Science Foundation of China (NSFC) under Grant Nos. 11925407 and 61927901. L.W. W was supported by the Director, Office of Science, the Office of Basic Energy Sciences (BES), Materials Sciences and Engineering (MSE) Division of the U.S. Department of Energy (DOE) through the theory of material (KC2301) program under Contract No. DEAC02-05CH11231.

**Author Contributions**

H.L. performed the TDDFT simulations and prepared the figures with the help of W.L. H.L. and W.L. conducted the analysis, discussion, and wrote the manuscript. J.L. and L.W. proposed the research project, established the project direction, and revised the paper with input from H.L and W.L. Z.S and Z.W contributed to the analysis and discussion of the data. S.L. provided the project infrastructure and supervised H.L.'s study. H.L. and W.L. contributed equally to this work.

# Supplementary Information for "Uniting the order and disorder dynamics in photoexcited VO$_2$"

Hao-Wen Liu[1,2§], Wen-Hao Liu[1,2§], Zhao-Jun Suo[1,2], Zhi Wang[1], Jun-Wei Luo[1,2]\*, Shu-Shen Li[1,2], and Lin-Wang Wang[3]\*

[1]*State Key Laboratory of Superlattices and Microstructures, Institute of Semiconductors, Chinese Academy of Sciences, Beijing 100083, China*

[2]*Center of Materials Science and Optoelectronics Engineering, University of Chinese Academy of Sciences, Beijing 100049, China*

[3]*Materials Science Division, Lawrence Berkeley National Laboratory, Berkeley, California 94720, United States*

[§]These authors contributed equally: Hao-Wen Liu, Wen-Hao Liu.
\*Email: jwluo@semi.ac.cn; lwwang@lbl.gov

# Supplementary Figures

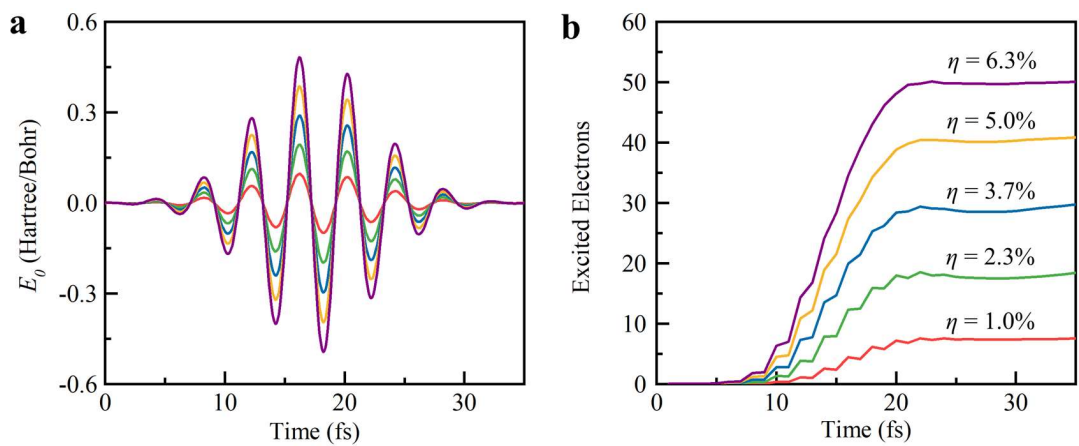

**Supplementary Figure 1. Laser-introduced excited electrons of $VO_2$. a,** The shape of the external electric field applied to the $VO_2$ with the laser strength $E_0$ = 0.1 to 0.5 V/Å. Similar lasers have been widely used experimentally to induce the phase transition of $VO_2$. The photon energy is 1.55 eV, corresponding to 800nm wavelength. The laser field value reaches the minimum strength at time $t_0$ = 35 fs. **b,** The number and percentage of excited electrons (η) upon photoexcitation from the valence bands to conduction bands under the different laser pulses.

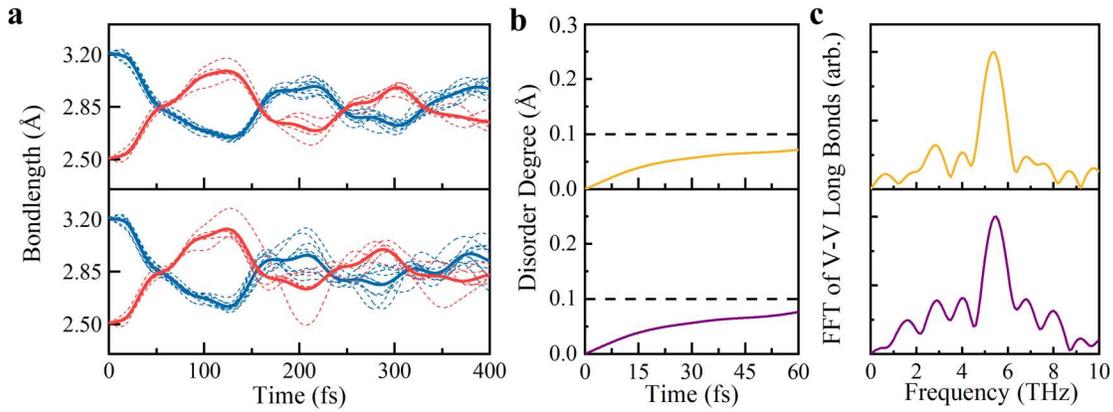

**Supplementary Figure 2. Photoinduced ultrafast phase transition in high electronic excitations. a,** The bond-length evolution under the different laser pumping. The blue and red dotted lines represent the bond lengths of each long and short bond, respectively. The blue and red solid lines represent the average bond length of long and short bonds. **b,** The disorder degree evolutions at different excitations before the phase transition time τ. **c,** Phonon modes of V atoms during the phase transition, which are obtained from the normalized fast Fourier transform (FFT) of the average bond length of V-V bonds.

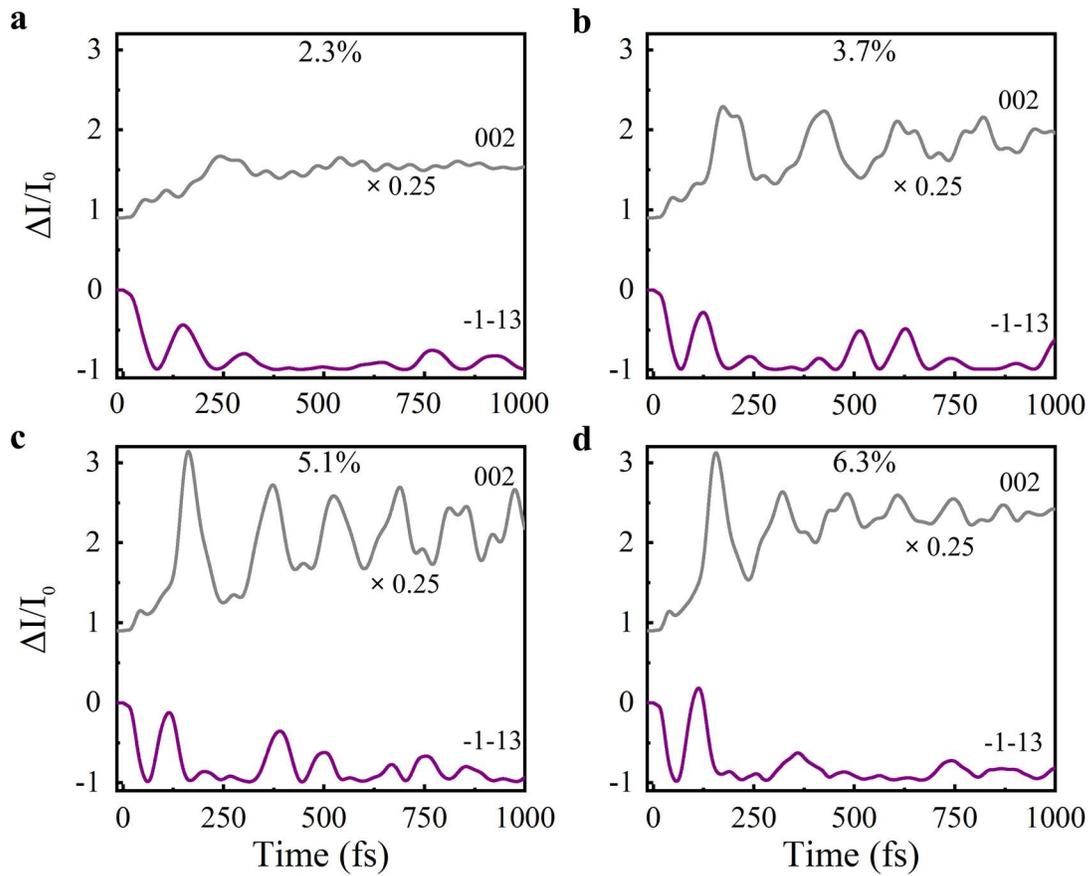

**Supplementary Figure 3. Time dependence of Bragg peaks for different photoexcitation. a-d,** Simulated Bragg peaks [(002)$_R$, (-1-13)$_{M1}$] as a function of time at different excitations, which are obtained through the structure factors. The results are the average diffraction intensities over eight unit cells within the 2×2×2 supercell structure.

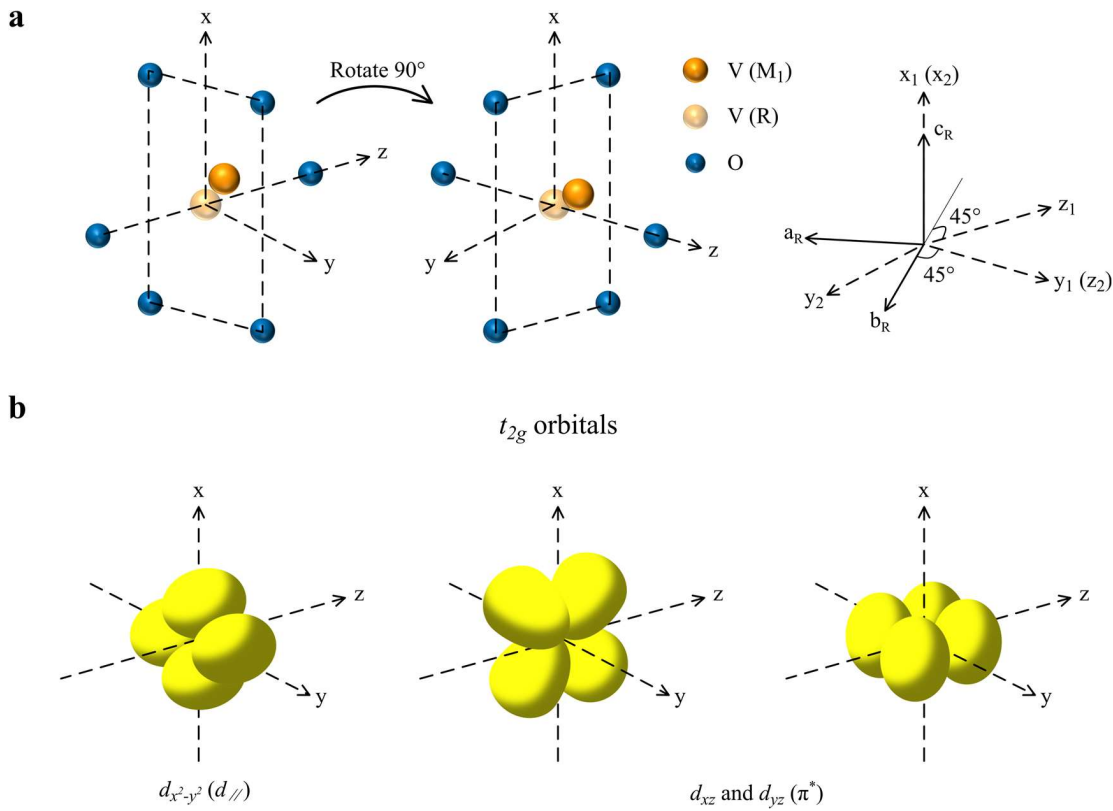

**Supplementary Figure 4. a,** In the rutile unit cell of $VO_2$, V atoms (imaginary orange atom) sites at the center of oxygen octahedrons, but in the monoclinic unit cell, V atoms (solid orange atom) derivates from the orthorhombic center. There is a 90° rotation angle between the two adjacent oxygen octahedrons, and a 45° rotation angle between the rutile crystallographic axes (solid arrows) and the geometric axes of the orbitals (dashed arrows). **b,** The schematic diagram of $t_{2g}$ ($d_{\parallel}$ forming valence electronic states and $\pi^*$ forming conduction electronic states) orbitals along the geometric axes ($x, y, z$) of the orbitals.

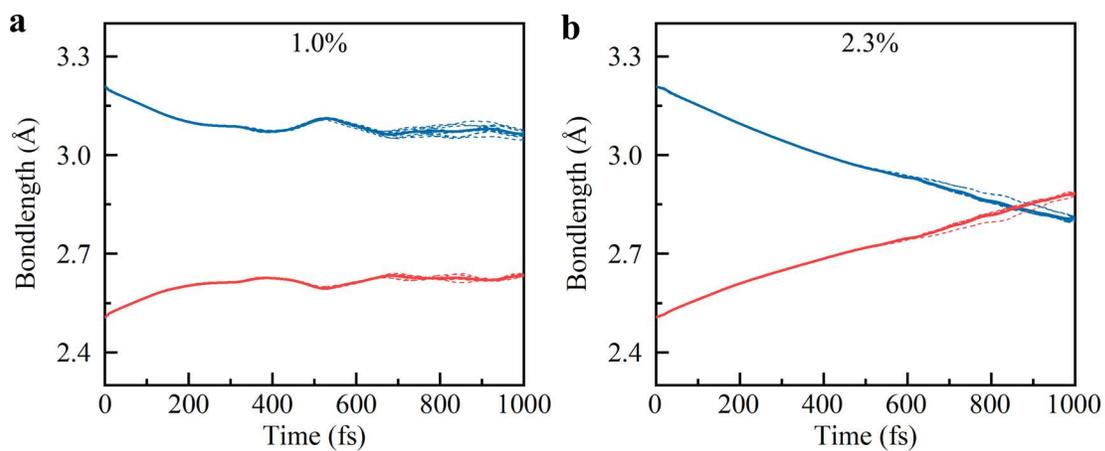

**Supplementary Figure 5. The bond-length evolution under the different laser pumping at the VO$_2$ system temperature ~ 1 K.** The blue and red lines represent the evolution of V-V long and short bonds, respectively. The dot and solid lines represent each pair and averaged V-V bond length.

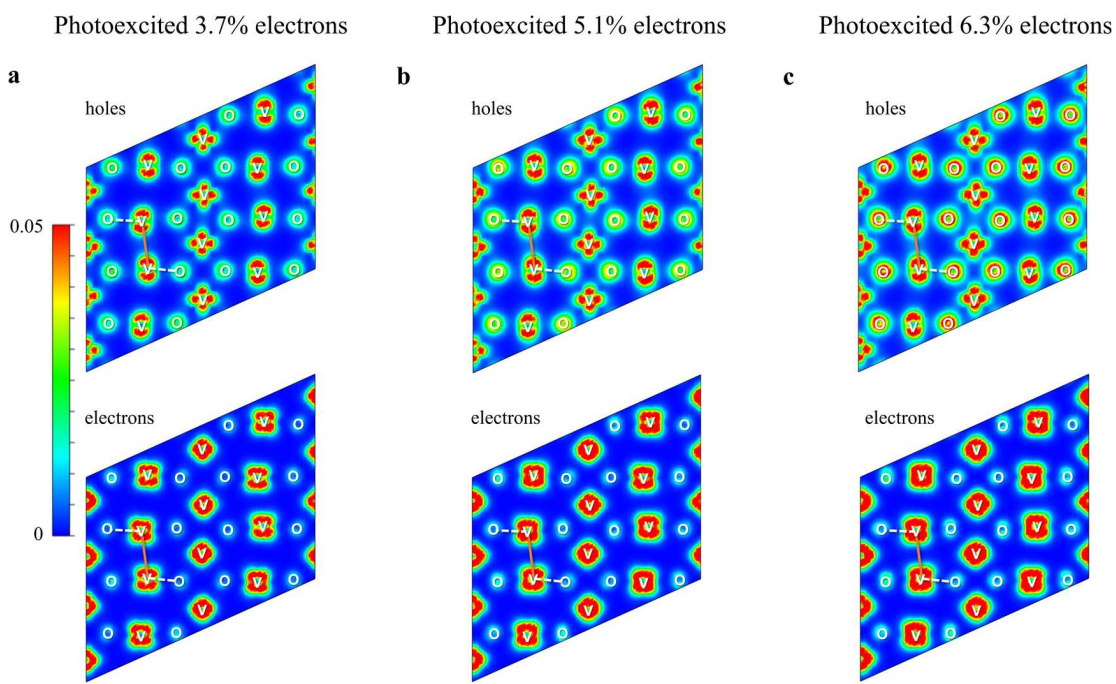

**Supplementary Figure 6.** Real-space distributions of photoexcited holes and electrons on ($0\bar{1}1$) plane at the end of the laser pulses (~35 fs) for the case of photoexcited 3.7%, 5.1%, and 6.3% valence electrons, respectively. As the number of photoexcited carriers increases, the holes near the O atoms increase significantly.